\documentclass[aip,rsi,reprint,graphicx]{revtex4-1} 
\usepackage{graphicx}
\usepackage[dvips, bookmarks, colorlinks=true]{hyperref}

\begin{document}
\title[The vectorial photoelectric effect under solar irradiance]{The vectorial photoelectric effect under solar irradiance and its application to sun sensing}
\author{Gerald Hechenblaikner and Tobias Ziegler}
\address{Airbus Defence and Space, 88039 Friedrichshafen, Germany}

\begin{abstract}
Sun sensors are an integral part of the attitude and orbit control system onboard almost any spacecraft. While the majority of standard analogue sun sensors is based on photo-detectors which produce photo-currents proportional to the cosine of the incidence angle (cosine detectors), we propose an alternative scheme where the vectorial photoelectric effect is exploited to achieve a higher sensitivity of the sensed photo-current to the incidence angle. The vectorial photo-effect is investigated in detail for metal cathode detectors in a space environment.  Besides long operational lifetimes without significant degradation, metal cathode detectors are insensitive to earth albedo, which may significantly reduce the errors affecting attitude measurements in low earth orbits. Sensitivity curves are calculated and trade-offs performed with the aim of optimizing the sensitivity whilst also providing currents sufficient for detection.  Simple applications and detector configurations are also discussed and compared to existing designs.
\end{abstract}

\maketitle

\section{Introduction}
\subsection{Sun Sensors}
Sun sensors are widely used as a part of spacecraft attitude determination subsystems to provide a measurement of the sun vector in spacecraft coordinates. For earth-pointing spacecraft, sun vector information may be used to complement the pitch and roll axes as obtained from horizon sensors with yaw axis measurements \cite{larson1992space}. Sun vector information might also be needed directly after launcher separation, to command the satellite into a sun acquisition mode where the solar panels are then deployed in order to establish electric power generation to allow spacecraft checkout operations\cite{friend2008orbital}. Because sun sensors are simple and typically have a wide field of view, they are often used during safe mode operation, where in case of anomalies the spacecraft is autonomously oriented in a thermally safe attitude where power supply via the solar panels is guaranteed. The sun vector information is often used to align the solar panel normal with the sun vector \cite{linder1992guidance}. In modern spacecraft the reference information to the attitude control system is typically provided by star trackers \cite{liebe1995star}. Some of these devices require sun sensor information to narrow down possibilities for spacecraft attitude and providing an initial scenario.
Usually, several sun sensors are accommodated on the spacecraft such that they cover the total celestial sphere. In failure cases, when the control system has lost attitude knowledge, the sun vector is directly used as input to the safe mode control system. When only a fractional area of the celestial sphere is covered, an autonomous recovery operation is used to rotate the spacecraft until the sun sensor has acquired the sun in its field of view\cite{doodydeep}.
Satellites operating in a geostationary orbit (GEO) are usually launched into a geosynchronous transfer orbit from where they are transferred to their final GEO position via apogee boost motor (ABM) firings. After ABM firings, the sun sensor information is used to command the satellite into a sun-pointing cruise mode to keep the batteries charged \cite{sidi2000spacecraft}.

Mainly two categories of suns sensors exist which consist either of digital or analog detectors \cite{wertz1978spacecraft}. Both sensor types mainly detect the visible part of the sun spectrum and are therefore sensitive to the albedo of nearby planets which might cause errors in the sun vector determination.
 Analog sun sensors are simple, lightweight, and inexpensive and are therefore often used in modern attitude control systems. The working principle of such sensors is based on the detection of the incident solar flux on one or multiple photo-diodes. The resulting photo-currents scale proportional to the angle between the direction of the incident light and the sensor surface normal, resulting in a cosine-dependence of the photo-currents from which the two sun angles can be derived. Commercially available satellites often use analog sun sensors. They are usually mounted on the solar panels of the satellite and provide a coarse yaw measurement.

In this paper we propose to use sun sensors based on metal photo-cathode detectors which offer several advantages compared to the conventional photo-diode based detectors\cite{ziegler2014sunsensor}. On the one hand, metal cathodes are the most robust of all photo-emitters\cite{dowell2010cathode} and offer many years of operational lifetime without significant degradation effects, which is particularly important in a hostile space environment.
On the other hand, the high work functions of most metals require light in the deep UV for photo-emission. The earth albedo (the sunlight reflected from earth) is close to zero for wavelengths below 240 nm\cite{frederick1987ultraviolet,feister1995spectral}, thus making a metal cathode based sun sensor essentially insensitive to earth albedo. This contrasts with conventional sun sensors which are sensitive to visible light, where the albedo has an average value of more than 30\%. This obviously strongly perturbs the conventional sun sensor and causes errors in the sensor readings on low earth orbits, which constitutes a major drawback that does not affect metal cathodes.
Finally, a phenomenon known as {\it vectorial photoelectric effect (VPE)} leads to strong variations of photo-currents with the angle of the incident light. This effect may be exploited in metal cathodes to obtain high sensitivity to small angular changes and robustness against variations of gain and light intensity in certain detector configurations.

\subsection{Spectral irradiance on a sensor}
\begin{figure}
\begin{center}
\begin{tabular}{c}
\includegraphics[width=7.0cm]{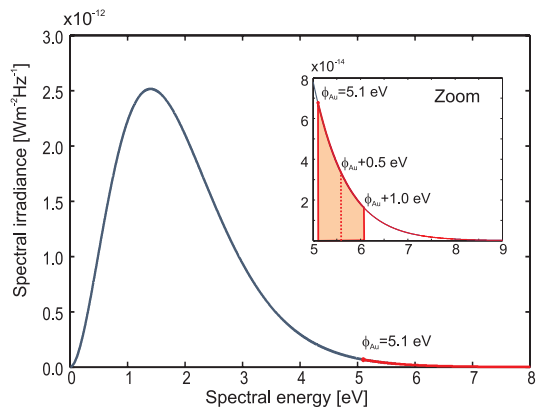}
\end{tabular}
\end{center}
\caption[example]
{\label{fig::solar_spectrum} The solar spectrum at a distance of one astronomic unit from the sun. The insert gives a zoom of the spectral range above the work-function of gold (5.1 eV).}
\end{figure}

The spectral irradiance $E_{e\nu}$ as described by Planck's law defines the energy radiated by an ideal black-body per unit area, time and frequency within a frequency interval $(\nu,\nu+d\nu)$.  The sun is well modeled as such a black-body with a temperature of 5780 K. We obtain the spectral irradiance on an object close to earth, e.g. a space vehicle in an earth orbit, by scaling Planck's law with an attenuation factor equal to the squared ratio between the radius of the sun $R_{\rm sun}$ and its distance to earth $R_{\rm au}$, as expressed in Eq.\ref{equ::bb_spectrum}:
\begin{equation}
E_{e\nu}d\nu=\frac{h\nu^3}{c^2}\frac{2\pi}{e^{\frac{h\nu}{k_BT}}-1}\left(\frac{R_{\rm sun}}{R_{\rm au}}\right)^2d\nu,
\label{equ::bb_spectrum}
\end{equation}
where $h$ denotes Planck's constant and $c$ is the speed of light. In the following discussions we assume that the detector surface is covered by a gold layer with a work function of $\phi_{Au}=5.1~{\rm eV}$, corresponding to wavelength of 243 nm, which defines a lower threshold energy for the detector. It is only photons with energies above this threshold which contribute to the photo-emission process.
Towards higher energies, the solar irradiance is truncated by a transmission window covering the detector surface. We assume the cutoff to be for wavelengths smaller than 200 nm (6.1 eV), which as discussed below will be beneficial for the sensitivity of the sun sensor. Figure \ref{fig::solar_spectrum} plots the full solar spectrum, while the insert zooms into the relevant region above 5.1 eV.
Integrating the solar spectrum over all frequencies, we obtain the total power from Stefan's law:
\begin{equation}
\int_0^{\infty}E_{e\nu}d\nu=\sigma T^4\left(R_{\rm sun}/R_{\rm au}\right)^2=1384~ {\rm Wm^{-2}},
\end{equation}
where $\sigma$ denotes Stefan's constant. If we constrain the integration to energies between the threshold and 1 eV above threshold (red shaded area in Fig.\ref{fig::solar_spectrum}), the intensity is found to be $8.7~{\rm Wm^{-2}}$ which amounts to only 0.63\% of the irradiance.

\section{The photo-electric effect}
\subsection{Photo-emission from the bulk}
Photo-electrons are emitted from a material surface, if the photon energy $h\nu$ of the incident light is higher than the material work function $\phi$. The quantum efficiency (QE) describes the number of photo-electrons generated per incident photon. It strongly depends on the frequency of the incident light.
In conventional literature and experiments, the photoelectric emission is mostly considered as coming from the material bulk. While a comprehensive description of the photo-emission process requires complex computations based on the specific atomic lattice structure and associated energy bands, the quantum efficiency is often well approximated by a simple expression referred to as Fowler's law\cite{fowler1931analysis}. This simple equation relates the quantum efficiency to the squared difference between the photon energy and the material work function, as given by Eq.\ref{equ::fowler_law}:
\begin{equation}
QE_{\rm bulk}=A_{PE}\left(1-R(\theta)\right)\left(h\nu-\phi_{\rm Au}\right)^2,
\label{equ::fowler_law}
\end{equation}
where $\theta$ denotes the angle of the incident light with respect to the surface normal, $R$ is the reflectivity of the surface, and $A_{PE}$ a material-dependent constant governing the overall strength of the specific emission process. Such a process is often described by a three-step model, originally developed by Spicer and coworkers\cite{spicer1958photoemissive,berglund1964photoemission,spicer2001modern}, in which excited electrons undergo inelastic collisions on their way to the surface where they must overcome a specific potential barrier.

\subsection{The vectorial photoelectric effect}
The quantum efficiency for bulk emission of photo-electrons, as described by Equation \ref{equ::fowler_law},  depends on the incidence angle solely through the reflectivity, but otherwise does not distinguish between the emission efficiency of electromagnetic field components parallel and normal to the material surface.
This contrasts the large variation of quantum efficiency with incidence angle predicted by the so-called {\it vectorial photoelectric effect (VPE)}, which was first described on the basis of a phenomenological model by Broudy\cite{broudy1971vectorial}. This effect assumes that normal components of the polarization vector of the incident light are much more likely to induce emission of an electron than parallel components. Therefore, in the case of p-polarized light where the polarization vector lies in the plane of incidence, this leads to emission currents strongly increasing with incidence angle. The effect has been observed on molybdenum\cite{juenker1964vectorial}, silicon\cite{broudy1971vectorial}, aluminum\cite{grepstad1976anisotropic}, and recently on copper\cite{pedersoli2005evidence,pedersoli2008surface}.
It is possible to rule out a dependency of  this effect on a specific crystalline geometry, as the vectorial photoelectric effect was found to be similarly strong in single-crystal as well as polycrystal surfaces\cite{pedersoli2005evidence}, where randomly oriented crystal domains would average out any symmetry-related contributions.

\subsection{A phenomenological approach to quantum efficiency}
\begin{figure}
\begin{center}
\begin{tabular}{c}
\includegraphics[width=7.0cm]{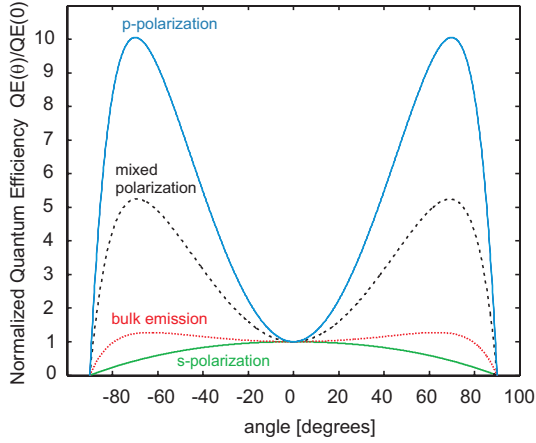}
\end{tabular}
\end{center}
\caption[example]
{\label{fig::cu_emission} The normalized quantum efficiencies for photoelectric emission from Cu with physical parameters taken from \cite{pedersoli2008surface}: work function $\phi_{Cu}=4.95\,{\rm eV}$, photon energy $h\nu=5.44\,{\rm eV}$ and cooperativity parameter $r=49$. In order of magnitude, from top to bottom, the calculated curves are given for p-polarized light (blue solid), mixed polarized light (black dashed), bulk emission (red dotted), and s-polarized light (green solid).}
\end{figure}

To obtain an expression for the quantum efficiency under consideration of polarization aspects, we use an ansatz where the total emission is written as the sum of contributions from polarization components parallel and normal to the surface\cite{broudy1971vectorial}. Each contribution is assumed to scale proportional to the respective absorbed energy, where $\epsilon_{\parallel}$ and $\epsilon_{\perp}$ denote the energy absorbed from the parallel and normal component, respectively. As we are interested in the variation of quantum efficiency with incidence angle, we normalize the total emission for a given angle $\theta$ by the emission at normal incidence ($\theta=0$). Considering that the emission efficiency due to polarization components normal to the surface is higher than for parallel components, we include a scaling factor r for the normal components, which we shall refer to as {\it cooperativity parameter} henceforth.
The total quantum efficiency is then written as:
\begin{eqnarray}
\frac{QE(\theta)}{QE(0)}&=&\frac{\epsilon_{\parallel}(\theta)}{\epsilon_{tot}(0)}+r\frac{\epsilon_{\perp}(\theta)}{\epsilon_{tot}(0)}\nonumber\\
&=&\frac{\epsilon_s(\theta)+\epsilon_{p\parallel}(\theta)}{\epsilon_s(0)+\epsilon_{p\parallel}(0)}
+r \frac{\epsilon_{p\perp}(\theta)}{\epsilon_s(0)+\epsilon_{p\parallel}(0)},
\label{equ::total_efficiency}
\end{eqnarray}
In the second line we have broken down $\epsilon_{\parallel}$ into one contribution from s-polarized light and a second contribution from the component of p-polarized light which is parallel to the surface:
$\epsilon_{\parallel}=\epsilon_s+\epsilon_{p\parallel}$. The energy absorbed from polarization components normal to the material surface, $\epsilon_{\perp}$, is exclusively given by the normal component of
p-polarized light: $\epsilon_{\perp}=\epsilon_{p\perp}$.
In case the incident light is polarized, we obtain from Equation \ref{equ::total_efficiency} expressions for the quantum efficiencies of s- and p-polarized light, respectively:
\begin{eqnarray}
\frac{QE_s(\theta)}{QE_s(0)}&=&\frac{\epsilon_s(\theta)}{\epsilon_s(0)}\\
\frac{QE_p(\theta)}{QE_p(0)}&=&\frac{\epsilon_{p\parallel}(\theta)}{\epsilon_{p\parallel}(0)}+r\frac{\epsilon_{p\perp}(\theta)}{\epsilon_{p\parallel}(0)}.\nonumber\\
\label{equ::efficiency_polarization}
\end{eqnarray}
In order to evaluate the quantum efficiencies, we must find the respective absorbed energies. The latter are found from expressions for the amplitudes of the electromagnetic radiation just inside the surface optical transition, as derived by Fan\cite{fan1945theory}.
\begin{eqnarray}
\frac{E_s}{E_i}&=&\frac{2\cos\theta}{s+\cos\theta}\nonumber\\
\frac{E_{p\parallel}}{E_i}&=&\frac{2s\cos\theta}{s+s_0^2\cos\theta}\nonumber\\
\frac{E_{p\perp}}{E_i}&=&\frac{2s\cos\theta}{s+s_0^2\cos\theta},
\label{equ::electric_amplitudes}
\end{eqnarray}
where we used the parameter $s=(s_0^2-\sin^2\theta)^{1/2}$. The parameter $s_0$ is the complex index of refraction, comprising the refraction index n and and the extinction coefficient $k$: $s_0=n+ik$.
The absorbed energy ratios are calculated from the electric field amplitudes as follows\cite{broudy1971vectorial}:
\begin{eqnarray}
\frac{\epsilon_{s}(\theta)}{\epsilon_{s}(0)}&=&\frac{1-R_s(\theta)}{1-R_s(0)}\nonumber\\
\frac{\epsilon_{p\perp}(\theta)}{\epsilon_{p\parallel}(0)}&=&\frac{1-R_p(\theta)}{1-R_p(0)}\frac{\left|E_{p\perp}(\theta)\right|^2}{\left|E_{p\perp}(\theta)\right|^2+\left|E_{p\parallel}(\theta)\right|^2}\nonumber\\
\frac{\epsilon_{p\parallel}(\theta)}{\epsilon_{p\parallel}(0)}&=&\frac{1-R_p(\theta)}{1-R_p(0)}\frac{\left|E_{p\parallel}(\theta)\right|^2}{\left|E_{p\perp}(\theta)\right|^2+\left|E_{p\parallel}(\theta)\right|^2}\nonumber\\
\label{equ::absorbed_energies}
\end{eqnarray}
Combining Equations \ref{equ::efficiency_polarization},\ref{equ::electric_amplitudes},\ref{equ::absorbed_energies}, we obtain the following expression for the quantum efficiency of p-polarized light:
\begin{eqnarray}
\frac{QE_p(\theta)}{QE_p(0)}&=& \frac{1-R_p(\theta)}{1-R_p(0)}\left[1+\left(r-1\right)\frac{\left|E_{p\perp}(\theta)\right|^2}{\left|E_{p\perp}(\theta)\right|^2+\left|E_{p\parallel}(\theta)\right|^2}\right]\nonumber\\
&=& \frac{1-R_p(\theta)}{1-R_p(0)}\left[1+\left(r-1\right)\frac{\sin^2\theta}{\sin^2\theta+\left|s\right|^2}\right].\nonumber\\
\label{equ::p_polarization}
\end{eqnarray}
Figure \ref{fig::cu_emission} plots the calculated quantum efficiencies due to the vectorial photoelectric effect for the various polarization types with parameters taken from Ref.\cite{pedersoli2008surface}. For comparison, we also plot the quantum efficiency for bulk emission (red dotted line), which features only a very weak dependence on angle due to the change of reflectivity compared to the strong variability for p-polarized light (blue solid line). The latter increases strongly towards angles of approximately $65^{\circ}$ before dropping rapidly to zero. The black dashed curve describes the quantum efficiency of mixed polarized light which is approximately half the quantum efficiency of p-polarized light.

\section{Photo-currents from solar irradiance}
\subsection{The variation of quantum efficiency with wavelength}
\label{sec::variation_with_nu}
In order to calculate the quantum efficiency over a large range of photon energies, it is necessary to know the value of certain physical parameters in the associated range of frequencies $\nu$. These parameters include the cooperativity parameter $r(\nu)$ and the complex index of refraction $s_0(\nu)$ from which the reflectivities for p- and s-polarized light $R_{p,s}(\theta,\nu)$ can be computed.

\begin{figure}
\begin{center}
\begin{tabular}{c}
\includegraphics[width=7.0cm]{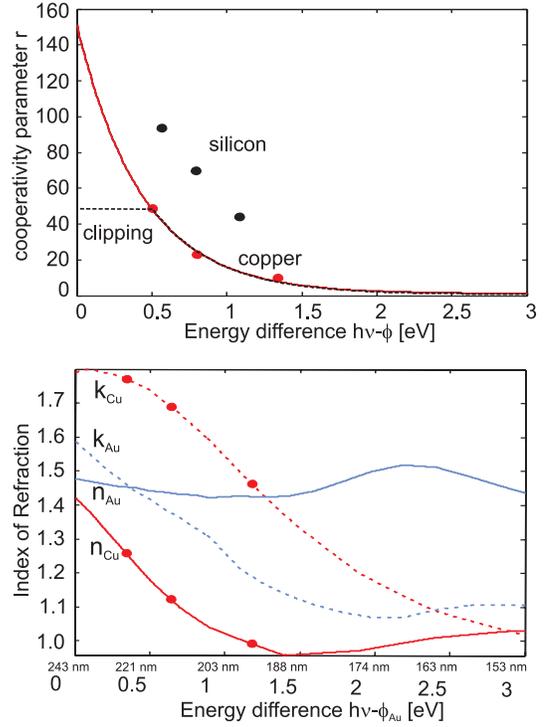}
\end{tabular}
\end{center}
\caption[example]
{\label{fig::nu_variations} The variation of basic physical parameters with photon energy. (top) The measured values of the cooperativity parameter in copper\cite{pedersoli2008surface} are plotted as red circles. An exponential fit is given by the red line. For a conservative worst case estimate, the fit is truncated at the highest measured value (black dotted line). For comparison, data measured for silicon are given by the black circles\cite{broudy1971vectorial}. (bottom) The values of the refractive index for copper (red lines) and gold (blue lines) are plotted against the difference between photon energy and the work function of gold ($h\nu-\phi_{\rm Au}$). The real part of the refractive index is drawn as a solid line, the imaginary part (extinction coefficient) as a dashed line.}
\end{figure}

Pedersoli showed in\cite{pedersoli2008surface} that the vectorial photoelectric effect is very strong close to the threshold and rapidly declines with increasing photon energies.
Whilst the data quoted in \cite{pedersoli2008surface} for copper and shown as red circles in Fig.\ref{fig::nu_variations}(top) provide the measured cooperativity parameter for three different photon energies, we require values $r(\nu)$ for all frequencies in the desired range. To this end, the data points are well fitted by an exponential function (red line). As a worst case estimate, in order to avoid overestimating the cooperativity parameter, we have also used a function that truncates the exponential function at the highest measured value (black dashed line) in comparative calculations of the yield currents in section \ref{sec:variation_with_angle}.

The variation of the complex index of refraction, which also enters the reflectivities $R_{p,s}(\theta,\nu)$, is plotted in  Fig.\ref{fig::nu_variations}(bottom) for copper and gold. The curves were obtained from interpolations of data given in \cite{palik1985handbook}. We see that both, refractive index $n$ and extinction coefficient $k$, change by less than a factor of 2 in the range of interest.

The proportionality constant $A_{PE}$ in Fowler's law is obtained by fitting the data for quantum efficiencies measured at various photon frequencies with a function that is quadratic in the frequency difference to threshold (see Eq. \ref{equ::fowler_law}). The data listed in \cite{pedersoli2008surface} for copper yield a value of $A_{PE}\approx 8.4\times 10^{-5}\,{\rm eV^{-2}}$. Measurements of the quantum efficiency of air-contaminated gold surfaces yielded a value of ${\rm QE=7\times10^{-5}}$ at normal incidence with light of $h\nu=4.9\,{\rm eV}$. Assuming that the work function was $\phi_{\rm Au}=4.2\,{\rm eV}$ (which is affected by a large uncertainty), this yields a value of $A_{\rm PE}\approx 21.6\times 10^{-5}\,{\rm eV}^{-2}$, which is higher by a factor of 2.5 than the measurements for copper. However, in order to remain conservative in our estimates of attainable photo-currents, we shall use the value of $A_{\rm PE}$ measured for copper in the following discussions, even though we propose to use a detector based on gold surface coatings.

\subsection{Spectral density of the photo-current}
\begin{figure}
\begin{center}
\begin{tabular}{c}
\includegraphics[width=8.0cm]{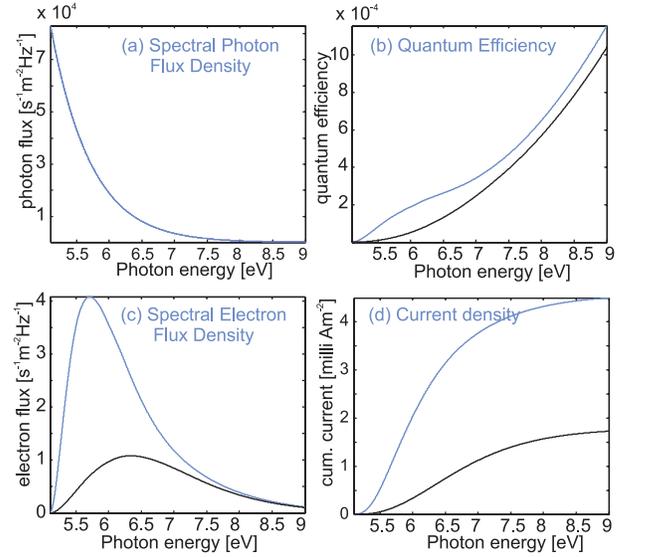}
\end{tabular}
\end{center}
\caption[example]
{\label{fig::emission_contributions} The variation of important physical quantities with photon energy for normal incidence (black solid line) and $45^{\circ}$ incidence (blue solid line). (a) incident photon flux. (b) total quantum efficiency. (c) flux of emitted photo-electrons. (d) integrated photo current density.}
\end{figure}

To obtain the spectral photon flux density, defined as the number of incident photons $n_{\nu}$ per unit of time, area, and frequency within a frequency interval $(\nu,\nu+d\nu)$, we divide the spectral irradiance $E_{e\nu}$ of Eq. \ref{equ::bb_spectrum} by the photon energy $h\nu$. The spectral photon flux density is plotted in Fig. \ref{fig::emission_contributions}a against the photon energy range of interest.

Considering that at zero incidence angle there is only bulk emission and inserting Eq. \ref{equ::fowler_law} into Eq. \ref{equ::p_polarization}, we obtain an expression for the increased quantum efficiency of p-polarized light in comparison to the quantum efficiency of bulk emission alone:
\begin{eqnarray}
QE_p(\theta,\nu)&=&A_{\rm PE}\left(h\nu-\phi_{\rm Au}\right)^2\left(1-R_p(\theta,\nu)\right)\nonumber\\
&\times& \left[1+\frac{\left(r(\nu)-1\right)\sin^2\theta}{\sin^2\theta+\left|s\right|^2}\right]\nonumber\\
&=& QE_{\rm bulk}(\theta,\nu)\left[1+\frac{\left(r(\nu)-1\right)\sin^2\theta}{\sin^2\theta+\left|s\right|^2}\right]]\nonumber\\
\label{equ::QE_p_final}
\end{eqnarray}
Equation \ref{equ::QE_p_final} demonstrates that the total emission comprises a part from the bulk and a part from the vectorial photoelectric effect which is enhanced by a factor of $(r-1)$ with respect to bulk emission. Figure \ref{fig::emission_contributions}b plots the quantum efficiency for p-polarized light at normal incidence (black curve) and at 45 deg incidence angle (blue curve). It becomes apparent that at energies close to threshold there is a vast enhancement of the efficiency due to the vectorial photoelectric effect, whilst at higher energies bulk emission dominates.

Having found expressions for the spectral photon flux density and the quantum efficiency in the relevant frequency range, it is possible to determine the photo-currents in the next step.
The spectral electron flux density, plotted in Fig. \ref{fig::emission_contributions}c, describes the number of electrons emitted per unit time, area and frequency within a frequency interval $(\nu,\nu+d\nu)$. It is given by the product of the spectral photon flux density of Fig.\ref{fig::emission_contributions}a and the quantum efficiency $QE$ of Fig.\ref{fig::emission_contributions}b. While the spectral photon flux density decreases with higher energies, the quantum efficiency increases. This leads to a peak in the spectral current density between 0.5 eV and 1.5 eV above threshold, depending on the actual incidence angle. Multiplying the spectral electron flux density with the elementary charge, we obtain the spectral photo-current density:
\begin{equation}
i_{\rm A\nu}=e\, n_{\nu}\,QE(\nu,\theta)=e\frac{E_{e\nu}}{h\nu}\,QE(\nu,\theta)
\end{equation}

Integrating $i_{A\nu}$ over the full spectral range from the threshold frequency $\nu_{\rm min}=\phi_{\rm Au}/h$ up to a variable cutoff frequency $\nu_{\rm max}$, we obtain the current density $i_{\rm A}$ at a certain incidence angle $\theta$:
\begin{eqnarray}
&&i_A(\theta,\nu_{\rm max})=\int_{\nu_{\rm min}}^{\nu_{\rm max}}e\frac{E_{e\nu}}{h\nu}\,QE_p(\theta, \nu)d\nu\nonumber\\
&&=e\left(\frac{R_{sun}}{R_{au}}\right)^2A_{\rm PE}\int_{\nu_{\rm min}}^{\nu_{\rm max}}d\nu\left\{\frac{2\pi\nu^2/c^2}{e^{\frac{h\nu}{k_BT}}-1}\left(h\nu-\phi_{\rm Au}\right)^2\right.\nonumber\\
&&\times\left.\left(1-R_p(\theta,\nu)\right)\left[1+\frac{\left(r(\nu)-1\right)\sin^2\theta}{\sin^2\theta+\left|s\right|^2}\right]\right\}\nonumber\\
\end{eqnarray}
Figure \ref{fig::emission_contributions}d plots the current density against the cutoff energy $h\nu_{\rm max}$ which is varied from 0 up to 4 eV above threshold. The current approaches an asymptotic limit of $4.5\times10^{-3}\,A/{\rm m}^2$ at 45 deg incidence and $1.8\times10^{-3}\,A/{\rm m}^2$ at normal incidence. For lower cutoff energies, the ratio between the photo-current at 45 deg incidence and the photo-current at normal incidence is very high, while it successively decreases towards higher energies and reaches a value of approximately 3 in the asymptotic limit. This gives an indication by how much the absolute photo-currents can be increased through the VPE effect as compared to bulk emission alone.

\subsection{Variation of photo-current with incidence angle}
\label{sec:variation_with_angle}
\begin{figure}
\begin{center}
\begin{tabular}{c}
\includegraphics[width=7.0cm]{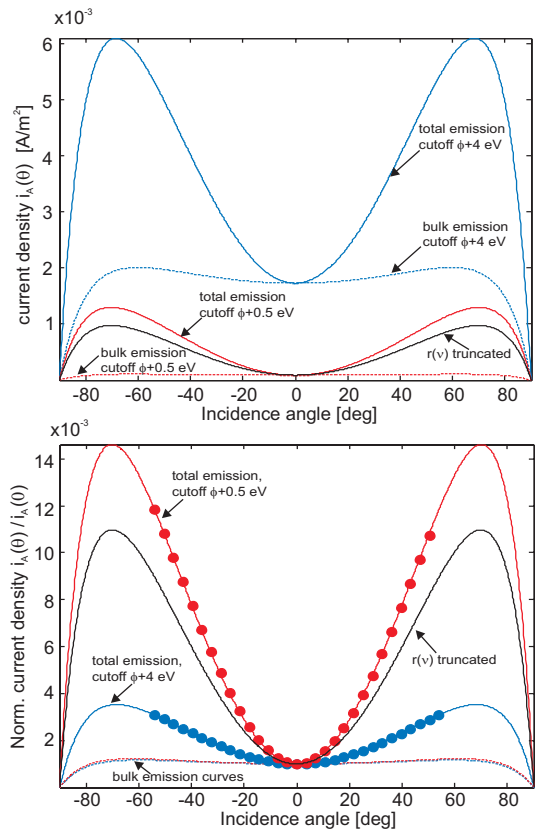}
\end{tabular}
\end{center}
\caption[example]
{\label{fig::gain_curves} The emission current density is plotted against angle of incidence in absolute terms (top) and relative to its value at normal incidence (bottom). The blue curves correspond to a photon cutoff energy at 4 eV above threshold, the red curves to a cutoff at 0.5 eV above threshold. The total emission is given by solid lines, the contribution from the bulk by the dashed lines. The black solid lines indicates the emission current for a worst case assumption of the cooperativity parameter and a cutoff at 0.5 eV above threshold. The solid circles describe an analytical approximation.}
\end{figure}

In the previous section we derived an expression for the spectral current density at a given incidence angle and plotted its cumulative integral against the cutoff energy for two specific angles in Fig.\ref{fig::emission_contributions}d. However, in order to quantify the sensitivity of the sensor to changes of incidence angle, it is useful to plot the photo-current over the full angular range for a given cutoff energy, as shown in Fig. \ref{fig::gain_curves} (top). The blue curves are calculated for a cutoff energy at 4 eV above threshold ($\lambda$=136 nm) where the current has approximately reached its asymptotic limit. While the solid lines refer to the total emission current, including the contribution from the vectorial photo-electric effect, the dashed lines refer to the bulk contribution alone. As expected, the total emission coincides with the bulk contribution at normal incidence but becomes significantly bigger for larger incidence angles where the VPE effect dominates. Peak values of the current density are expected to be around $6\times10^{-3}\,{\rm A/m^2}$ at 60 deg incidence angle.

The red curves were calculated for an energy cutoff at 0.5 eV above threshold ($\lambda$=221 nm), where the current density peaks at $1.3\times10^{-3}\,{\rm A/m^2}$, which is significantly lower than the limit achieved in the asymptotic energy cutoff limit. However, the advantage of increased sensitivity more than outweighs the drawback of reduced absolute currents.
This is demonstrated in Fig. \ref{fig::gain_curves} (bottom), where the emission currents of Fig. \ref{fig::gain_curves} (top), normalized by their values at normal incidence, are plotted against the incidence angle $\theta$. It shows that the emission current with a cutoff in the asymptotic limit increases by a factor of only 3.5 between 0 and 60 deg, while the factor is more than 14 for the emission current with cutoff at 0.5 eV above threshold. If the cutoff is chosen higher, at 1 eV above threshold, the current peaks at a higher value of $3.5\times10^{-3}\,{\rm A/m^2}$ after increasing by a factor of only 8.5 with respect to normal incidence. Such a decrease in angular sensitivity in relative terms is expected for higher cutoff energies, as the VPE effect is only strong for photon energies close to threshold and quickly fades for higher energies. This was discussed in section \ref{sec::variation_with_nu} and was summarized by the plot for the cooperativity parameter $r(\nu)$ in Fig.\ref{fig::nu_variations}. We noted that the extrapolation of the cooperativity parameter $r(\nu)$ from the measurement data towards lower photon energies with an exponential function might overestimate its values in that range. In order to obtain a worst case estimate, we introduced a cutoff of $r(\nu)=50$  at the last measured value approximately 0.5 eV above threshold (indicated by the black dotted line in Fig. \ref{fig::nu_variations}). The resulting density of the total emission current is given by the black solid line in Figs.\ref{fig::gain_curves} (top) and (bottom). Fortunately, we find that even in this worst case scenario the emission current is close to the one obtained for an exponential trend of the cooperativity parameter (red solid line). This can be explained by the strong increase of the spectral density of the electron flux between 0 and 1 eV above threshold (see Fig.\ref{fig::emission_contributions}c), indicating that the photons of higher energy (and therefore lower cooperativity) dominate the total emission current.
The total emission currents can be approximated by the following relation in the range from 0 to approximately 60 deg incidence angle:
\begin{equation}
i_A(\theta)=i_A(0)\left(1+B\sin^2\theta\right),
\label{equ:sin_square_fit}
\end{equation}
where $B$ is a parameter describing the sensitivity. The functional appearance of this analytical approximation is motivated by comparison to Eq.\ref{equ::QE_p_final} for small angles $\theta$.
The solid circles of Fig.\ref{fig::gain_curves} (bottom) represent a fit to the total emission current based on Eq.\ref{equ:sin_square_fit}, which shows that it is indeed a very good approximation. For the total emission current with cutoff at $0.5\,{\rm eV}$ above threshold (red curve) we find $B=16$.

\section{Application to a sun sensor}
In this section we shall briefly discuss how metal cathode detectors can be used in various configurations as a sun sensor. While our investigations only cursorily touch the subject matter without details of the implementation and proof of its feasibility, we aim to bring forward some general ideas that may be applied and elaborated in similar schemes.
\subsection{Schematic of a VPE detector}
\begin{figure}
\begin{center}
\begin{tabular}{c}
\includegraphics[width=7.0cm]{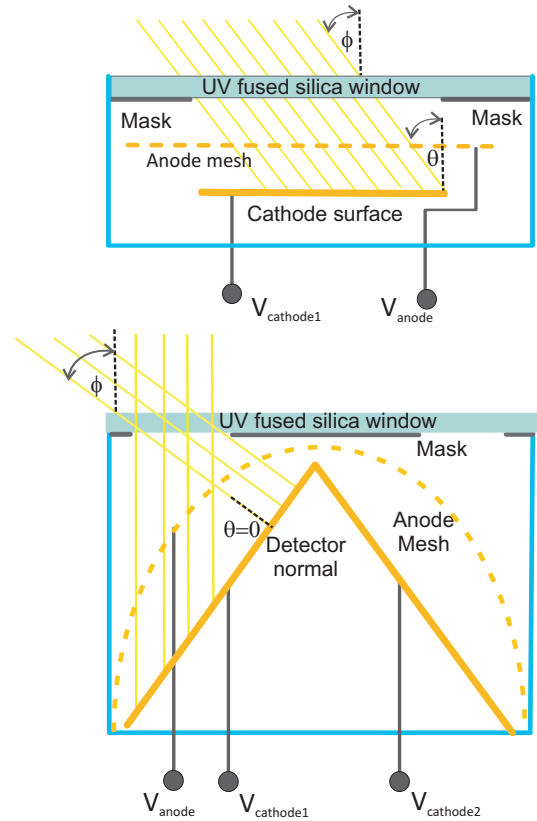}
\end{tabular}
\end{center}
\caption[example]
{\label{fig::detector_designs} Examples for detector designs. (top) A basic VPE detector has a single cathode surface. The incidence angle on the aperture $\phi$ equals the incidence angle on the detector surface $\theta$. (bottom) A VPE detector based on metal cathodes which are tilted with respect to the input aperture. }
\end{figure}

An example for a basic metal cathode detector is given in Fig.\ref{fig::detector_designs} (top). The solar irradiance is transmitted through the input aperture covered by UV fused silica glass, which may also include a transmission filter and polarizer, before it is incident on the metal cathode surface.
 Typically, UV fused silica windows (UVFS) transmit light down to wavelengths around 200  nm ($6.1\,{\rm eV}$). Alternatively, windows may be made from magnesium fluoride or sapphire which also transmit in the required wavelength range. In order to cut-off transmission for wavelengths below 220 nm ($0.5\, {\rm eV}$ above threshold), a reflective coating could be applied on the window that is optimized for high reflectivity at lower wavelengths. Including a polarizer, e.g. a dielectric nanowire polarizer, selects a sensitive incidence plane. Consequently, the detector is primarily sensitive to angular changes of the incident p-polarized light in the plane spanned by the polarizer axis and the detector surface normal. If no polarizer is used, the incident light from the sun has mixed polarization, which reduces the sensitivity by a factor of approximately 2 compared to p-polarized light (see black dashed line in Fig.\ref{fig::cu_emission}). The detector may still be operated in such a configuration, but this comes at the expense of reduced angular sensitivity.

 A voltage difference of approximately 15 V is applied between the cathode and the anode mesh which receives the emitted photo-electrons. In order to determine the photo-currents, it is preferable to measure the current that is sourced by the metal cathode rather than the current that is sunk by the collecting anode. These two currents may differ, if the anode also collects electrons from other sources than the metal cathode, which may happen in a free space arrangement where the electrodes are directly exposed to space. Although such an arrangement is highly susceptible to surface contamination, it might be possible to exploit the natural occurrence of a UV-cleaning process in space. It was shown by Vig \cite{vig1985uv,vig1986uv,reinhardt2008handbook} that organically contaminated gold surfaces experience a cleaning process when interacting with UV-light in combination with oxygen. This process relies on the decomposition of organic molecules, in particular hydro-carbons, by the UV-light which breaks up the contaminant molecule bonds. Similarly, it has also been pointed out that if a clean surface is stored under UV-radiation, the surface cleanliness may be maintained indefinitely\cite{sowell1974surface}. However, considering the ease of ground testing and in order to avoid any kind of contamination, the vacuum sealed detector configuration is the recommended baseline.

\subsection{Geometric detector configurations}
So far we have assumed a constant photon flux from the sun onto the detector surface without considering the impact of geometry. In reality, the actual photo-current also depends on the orientation of the input aperture with respect to the incident light, which is described by a cosine projection factor.
The red curve of Fig.\ref{fig::gain_curves} depicts the emission current density for a constant photon flux. In order to obtain the detector current, the current density $i_A$ is multiplied by the area $A_{\rm ap}$ of the input aperture and by the cosine of the angle $\phi$ between the aperture normal and the direction of the incident light:
\begin{equation}
I(\theta)=i_A(\theta)A_{\rm ap}\cos\phi
\label{equ::cosine_law}
\end{equation}
Note that the $\cos\phi$ factor of Eq.\ref{equ::cosine_law} is the geometric projection factor on which the fundamental working principle of a cosine detector is based.
If the input aperture plane is parallel to the detector surface, the angles $\phi$ and $\theta$ coincide, see Fig.\ref{fig::detector_designs} (top), and the detector current is given by Eq.\ref{equ::parallel_geometry}:
\begin{equation}
I(\theta)=i_A(0)A_{\rm ap}\left(1+B\sin^2\theta\right)\cos\theta
\label{equ::parallel_geometry}
\end{equation}
For the basic detector of Fig.\ref{fig::detector_designs} (top), the geometric projection factor decreases with increasing incidence angle while the quantum efficiency increases.
However, if the detector surface is tilted with respect to the input aperture, as shown in Fig.\ref{fig::detector_designs} (bottom), this allows for both, geometric projection factor and quantum efficiency, to increase with incidence angle, which leads to higher sensitivity. Assuming an angle of $60^{\circ}$ between aperture and detector surface, the detector current is given by:
\begin{equation}
I(\theta)=i_A(0)A_{\rm ap}\left(1+B\sin^2\theta\right)\cos\left(\theta+60^{\circ}\right).
\label{equ::tilted_geometry}
\end{equation}
One scenario in Figure \ref{fig::detector_designs} (bottom) depicts light at a large incidence angle $\phi$ with respect to the input aperture, leading to a small photon flux arriving at normal incidence ($\theta=0$) on the detector surface, which generates the minimum photo-current. The second scenario depicts light at normal incidence on the input aperture, leading to a large photon flux at a large incidence angle (close to $60^{\circ}$) on the detector surface, which generates the maximal current. Different configurations are possible where currents may be added or subtracted from another. It requires more detailed analyses and experimental verification to determine which schemes might potentially be suitable for application as a sun sensor in space.

\subsection{Sensitivity and robustness}
At the end of section \ref{sec:variation_with_angle} we noted that the VPE detector has a sensitivity scaling with the parameter B of Eq.\ref{equ:sin_square_fit}.
In this section we shall discuss sensitivity in more detail and look at the robustness of VPE detector readings against intensity fluctuations of the incident light. For simplicity, we shall consider a constant photon flux and omit the geometric projection factor $\cos\phi$, as it depends on the specific choice of detector geometry.

The sensitivity is affected by the incremental change of photo-current due to a small change of the solar incidence angle. At any particular instant, when the sun light is incident on the detector surface at an angle $\theta$, we may either consider the current change relative to a fixed reference value, e.g. the current at normal incidence $I(0)$, or relative to its instantaneous value $I(\theta)$.
These two cases define the sensitivity parameters $K_{s0}$ and $K_{s\theta}$, respectively, which are calculated in Tab.\ref{tab:sensitivities} for a VPE and a cosine detector.
\begin{table}
\caption{\label{tab:sensitivities}The sensitivity parameters $K_{s0}$ and $K_{s\theta}$  are calculated for VPE and cosine detectors, respectively.}
\begin{tabular}{cccc}
  Notation & Definition & VPE det.& Cosine det.  \\
  \hline
  $K_{s0}$ & $\frac{1}{I(0)}\frac{dI(\theta)}{d\theta}$ & $B\sin 2\theta$ & $-\sin\theta$  \\
  $K_{s\theta}$ & $\frac{1}{I(\theta)}\frac{dI(\theta)}{d\theta}$ & $\frac{B\sin 2\theta}{1+B\sin^2\theta}$ & $-\tan\theta$ \\
  \hline
\end{tabular}
\end{table}
We find from Tab.\ref{tab:sensitivities} that the parameter $K_{s0}$ for a conventional cosine detector is always smaller than 1 while it may go up to a value of $B$ for a VPE detector.

\begin{figure}
\begin{center}
\begin{tabular}{c}
\includegraphics[width=7.0cm]{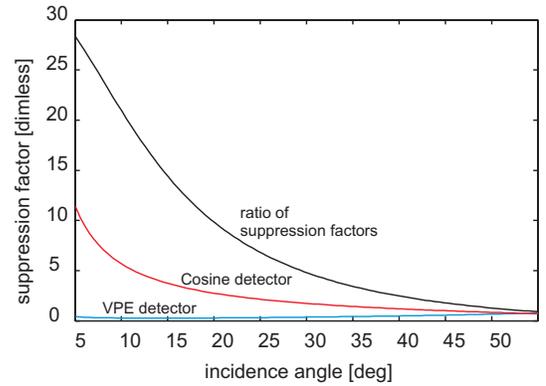}
\end{tabular}
\end{center}
\caption[example]
{\label{fig::gain_changes} The suppression factor for gain variations $|1/K_{s\theta}|$ is plotted for a cosine (red line) and a VPE (blue line) detector.
A value of $B=16$ was assumed for the VPE detector. The ratio between the two factors is given by the black curve.}
\end{figure}

If  angular changes are referenced to the instantaneous values, as described by the parameter $K_s$, one obtains a measure for relative sensitivity which is inversely proportional to the stability of the angular reading against intensity variations.
This can be seen from the ansatz of Equation \ref{equ:intensity_fluctuations}, where a small fractional change of the light intensity $\epsilon$ leads to a corresponding error $\Delta\theta$ in the measured angle of a VPE detector:
\begin{eqnarray}
I_0\left(1+\epsilon\right)\left(1+B\sin^2\theta\right)&=&I_0\left(1+B\sin^2\left(\theta+\Delta\theta\right)\right)\nonumber\\
&\Rightarrow&\Delta\theta=\frac{\epsilon}{K_{s\theta}},\nonumber\\
\label{equ:intensity_fluctuations}
\end{eqnarray}
where we used a Taylor expansion to first order in $\Delta\theta$ to arrive at the result of the second line.
This equation demonstrates that a high angular sensitivity reduces errors induced by fluctuations in the overall gain by a suppression factor $|1/K_{s\theta}|$. For comparison, this factor is plotted for a VPE (blue curve) and a cosine detector (red curve) in Fig.\ref{fig::gain_changes}. The ratio between the two suppression factors is also plotted (black curve), indicating that the a VPE detector exhibits greater robustness for angles up to approximately 50 deg.  Note that in order to mitigate this weakness of a conventional cosine detector, it is often operated around a reference point defined for a large angle of incidence (instead of normal incidence), which makes up for the reduced sensitivity compared to a VPE detector. However, such a configuration comes at the expense of a reduced angular range and a limited field of view.

\section{Conclusions}
We have investigated the photo-currents generated by metal cathode detectors under solar irradiance in space which typically range from $100$ to $600\,\mu{\rm A/cm^2}$, depending on the incidence angle and the spectral cutoff frequency. The photo-currents vary strongly with incidence angle for metals exhibiting the vectorial photoelectric effect, which may be exploited for increased sensitivity to small angular changes compared to a mere projection of the incident solar flux according to a cosine law. As the vectorial photoelectric effect has its origins in emission from the material surface state, it is primarily sensitive to light close to the threshold frequency. Consequently, we studied how the angular sensitivity may be improved, albeit at the expense of smaller emission currents, by reducing the cutoff frequency of the solar spectrum.
We propose to use such detectors in future sun sensors, which could significantly reduce aging and degradation effects due to the inherent robustness of metal cathodes. Additionally, such sensors would be
insensitive to earth albedo as the typical work functions of metals are in the deep UV where hardly any light is reflected from the earth. This makes them attractive for use in low earth orbit missions. Various detector configurations are discussed and the differences to cosine detectors are analyzed. While we presented some preliminary assessments, more detailed studies accompanied by experiments need to be performed to better judge the feasibility of the proposed concepts.

\section*{Acknowledgements}
 A patent application (14000559.6) containing basic schematics on the use of VPE detectors in a sun sensor has been filed to the European Patent Office (EPO)\cite{ziegler2014sunsensor}. The authors are grateful to Patrick Bergner, Nico Brandt, and Michael Kersten for fruitful discussions. They would also like to thank Michael Kersten for a critical review of the manuscript with suggestions for improvement and Georg Willich for his help and support.

{\it Author contributions}: The idea of using a VPE sensitive detector for sun sensing comes from T. Ziegler and was subsequently analyzed by G. Hechenblaikner. Both authors contributed equally to the paper.

\section*{References}

\end{document}